# Large and robust electrical spin injection into GaAs at zero magnetic field using an ultrathin CoFeB/MgO injector

S. Liang[1], T.T. Zhang[2], P. Barate[2], J. Frougier[3], M. Vidal[2], P. Renucci[2], B. Xu[4], H. Jaffrès[3], J.M. George[3], X.Devaux[1], M. Hehn[1], X. Marie[2], S. Mangin[1], H. Yang[5], A. Hallal[5], M. Chshiev[5], T. Amand[2], H. Liu[6], D. Liu[6], X. Han[6], Z. Wang[4], Y. Lu[1*]

[1]*Institut Jean Lamour, UMR 7198, CNRS-Nancy Université, BP 239, 54506 Vandoeuvre, France*

[2]*Université de Toulouse, INSA-CNRS-UPS, LPCNO, 135 avenue de Rangueil, 31077 Toulouse, France*

[3] *Unité Mixte de Physique CNRS/Thales and Université Paris-Sud 11, 1 avenue A. Fresnel, 91767 Palaiseau, France*

[4]*Key Laboratory of Semiconductor Materials Science, Institute of Semiconductors, Chinese Academy of Sciences, P. O. Box 912, Beijing 100083, China*

[5] *SPINTEC, UMR 8191, CEA-INAC/CNRS/UJF-Grenoble 1/G-INP, 38054 Grenoble*

[6]*Beijing National Laboratory for Condensed Matter Physics, Institute of Physics, Chinese Academy of Sciences, P.O. Box 603, Beijing 100190, China*

*Corresponding author: yuan.lu@univ-lorraine.fr

**Abstract**

We demonstrate a large electrical spin injection into GaAs at zero magnetic field thanks to an ultrathin perpendicularly magnetized CoFeB contact of a few atomic planes (1.2 nm). The spin-polarization of electrons injected into GaAs was examined by the circular polarization of electroluminescence from a Spin Light Emitting Diode with embedded InGaAs/GaAs quantum wells. The electroluminescence polarization as a function of the magnetic field closely traces the out-of-plane magnetization of the CoFeB/MgO injector. A circular polarization degree of the emitted light as large as 20% at 25 K is achieved at zero magnetic field. Moreover the electroluminescence circular polarization is still about 8% at room temperature.

Since the discovery of an efficient transfer of a solid-state information stored within ferromagnetic materials into circularly polarized photons emitted by spin light emitting diode (spin-LED)[1,2] via carrier-photon angular momentum conversion, several advanced semiconductor technologies have been proposed. Potential devices ranging from memory element with optical readout and optical transport of spin information[3], advanced optical switches[4], circularly polarized single photon emitters for quantum cryptography[5] to chiral analysis[6] and 3-dimensional display screens[7] have been anticipated. According to the optical selection rules[8] in quantum well (QW)-based spin LEDs, conventional spin-injector with in-plane magnetization[9-18] cannot satisfy the practical application because a strong external magnetic field in the range of up to a few Tesla is required to rotate the magnetization into a perpendicular direction. A prerequisite to obtain optimized device functionalities is to promote a robust *perpendicular* magnetic anisotropy (PMA) medium up to room temperature (RT) to be used as a solid-state ferromagnetic (FM) injector electrode. Good candidates are systems including alternated planes of 3$d$/4$f$ Fe/Tb[19,20], 3$d$/5$d$ Co/Pt [21,22], or 3$d$/3$d$ Co/Ni multilayers[23]. However, these systems generally suffer from the requirement of a minimum thickness (generally several units of bilayers) grown on a thin oxide layer, which is used as a tunneling barrier to circumvent the conductivity mismatch between metal and semiconductor[24]. The large thickness of injector results in a large absorption of light in the near infrared region, e.g. 95% light is absorbed for 40 nm Fe/Tb multilayers[19,20]. Moreover, in the case of spin-LEDs, there is also a requirement that the first FM atomic plane grown at the interface must possess a robust spin-polarization for an efficient spin-injection, which is hardly attainable due to the chemical inter-diffusion or intermixing in the multilayer systems[22]. Therefore, up to now, the circular polarization ($P_C$) of emitted light was still limited to 3-4% at remanence[20,22,25]. A series of recent theoretical investigations have proposed that the Fe(Co)/MgO interface itself could provide PMA in the range of magnitude of 1mJ/m$^2$, sufficient to reorient the magnetization along the out-of-plane direction[26,27]. This PMA property has been put forward in the case of CoFeB/MgO/CoFeB p-MTJs grown on SiO$_2$ substrate used for spin transfer torque (STT) operations in MRAM technologies (STT-MRAM)[28,29]. These ultrathin CoFeB electrodes are very well suited for STT devices because they allow high tunneling magnetoresistance (TMR) ratio and display good thermal stability together with low switching current density. In this Letter, we demonstrate the occurrence of such PMA functionality on semiconductor heterostructures with III-V based spin-LEDs by integrating CoFeB/MgO perpendicular spin-injectors. Large values of electroluminescence circular polarization of 20% at 25K and 8% at 300K are measured under zero magnetic field.

We have investigated electrical spin injection from an ultrathin perpendicularly magnetized CoFeB contact (1.2 nm) into InGaAs/GaAs quantum wells (QWs) embedded spin-LED. The whole structure of sample is schematically shown in Fig. 1(a). The *p-i-n* LED device grown by molecular beam epitaxy (MBE) has the following structure sequence: *p*-GaAs:Zn (001) substrate ($p=2\times10^{19}$cm$^{-3}$) /500 nm *p*-GaAs:Be ($p=2\times10^{19}$cm$^{-3}$) /100 nm *p*- Al$_{0.3}$Ga$_{0.7}$As:Be ($p=2\times10^{19}$cm$^{-3}$) /100 nm *p*-Al$_{0.3}$Ga$_{0.7}$As:Be ($p=2\times10^{18}$cm$^{-3}$) /50 nm undoped Al$_{0.3}$Ga$_{0.7}$As /[15 nm undoped GaAs/ 8 nm undoped In$_{0.1}$Ga$_{0.9}$As QW]×3 /15 nm undoped GaAs/ 5 nm undoped Al$_{0.3}$Ga$_{0.7}$As /30 nm undoped GaAs /50 nm *n*-GaAs:Si ($n=1\times10^{16}$cm$^{-3}$). The surface was passivated with arsenic in the III-V MBE chamber. The intended design of the 3QWs structure for LED is to obtain strong electroluminescence intensity especially at room temperature. The sample was then transferred through the air into a magnetron sputtering-MBE interconnected system to grow the CoFeB/MgO spin-injector. The arsenic capping layer was firstly desorbed at 300°C by monitoring *in situ* reflection high energy electron diffraction (RHEED) patterns in the MBE chamber, and then the sample was transferred to the sputtering chamber to grow the spin-injector. The spin-injector grown at room temperature consists in a 2.5 nm MgO tunnel barrier and a thin Co$_{40}$Fe$_{40}$B$_{20}$ ferromagnetic layer (1.2 nm). Finally, 5nm Ta was deposited to prevent oxidation. 300 μm diameter circular mesas were then processed using standard UV photolithography and etching techniques. In the end, the processed wafers were cut into small pieces to perform rapid temperature annealing (RTA) in the temperature range 250-300°C for 3 minutes. The RTA procedure is a good way to promote PMA of CoFeB[30] while almost keeping no change to the LED optical characteristics. High-resolution transmission electron microscopy (HRTEM) study was performed using a JEOL ARM200 cold FEG gun working at 200kV. The magnetic property of CoFeB layer was measured by superconducting quantum interference device magnetometer (SQUID). For the polarization-resolved EL measurements, the spin-LED was placed into a Helmoltz-split magnetic coil providing a maximum magnetic field of 0.8T normal to the sample plane. The EL signal was detected in the Faraday geometry. The spectral resolution of the set-up is 2 meV (1.3 nm). The EL circular polarization $P_C$ was analyzed through a λ/4 wave plate and a linear analyzer. $P_C$ is defined as $P_C=(I^{\sigma+}-I^{\sigma-})/(I^{\sigma+}+I^{\sigma-})$ where $I^{\sigma+}$ and $I^{\sigma-}$ are the intensities of the right and left circularly polarized components of the luminescence, respectively.

The interface of the spin-injector consisting of a 1.2 nm CoFeB layer after RTA annealing at 300°C was investigated by HRTEM. The low magnification image [inset of Fig. 1(b)] reveals a good homogeneity and a very low roughness of MgO on GaAs. Moreover, the continuity of the 1.2 nm CoFeB layer can be also validated in this picture. From the high magnification image [Fig. 1(b)], a MgO (001) texture is clearly revealed with an abrupt interface to both the CoFeB and GaAs layers, whereas the interface of Ta/CoFeB whose quality is less important for PMA appears rather diffusive. By careful examination of the interface between CoFeB/MgO, we can depict some small crystalline grains of CoFeB indicating that the crystallization of CoFeB starts from the CoFeB/MgO interface[30]. As our best electroluminescence (EL) results are obtained at low temperature, Fig. 2(c) shows the in-plane and out-of-plane *M-H* curves at 30K for the perpendicular injector with $t_{CoFeB}$=1.2 nm. We can observe a clear perpendicular easy axis with out-of-plane coercivity $\mu_0H_C$=20 mT and in-plane saturation field $\mu_0H_K$=150 mT. The perpendicular anisotropy energy density $K_{eff}$ is then determined to be $4.6\times10^4$ J/m$^3$, which is lower than the one obtained on CoFeB p-MTJs ($2.1\times10^5$ J/m$^3$) [28].

We focus now on polarization-resolved EL measurements on a spin-LED after annealing at 300°C. A typical EL spectrum acquired at 25K under a bias of 2.30V is shown in the top of Fig. 2(a) for $\mu_0H$=0T. In this spectrum, we can observe a main peak located at about 873 nm corresponding to the heavy-hole exciton line, with a small shoulder at about 870 nm. The multi-peak feature could be attributed to a slight different indium concentration for the three InGaAs QWs as well as the possible bound exciton at low temperature[31,32]. The striking feature is that we can get a large difference of the EL intensities for right ($I^{\sigma+}$) and left ($I^{\sigma-}$) circularly polarized components at zero field. The EL circular polarization ($P_C$) can be determined from the main peak difference for $I^{\sigma+}$ and $I^{\sigma-}$ to be about 13%. To further confirm that this feature originates from the perpendicular spin-injector, we have measured the $P_C$ variation at different magnetic field. As shown in Fig. 2(b), $P_C$ exhibits a clear hysteresis loop feature with almost constant value around 13% at saturation and changing its sign rapidly at $\mu_0H$=±30mT. The hysteresis loop of $P_C$ fairly matches the SQUID hysteresis loop acquired at 30K on an unpatterned sample [Fig. 2(c)]. The slight difference in the coercivity could be due to a different measuring temperature in the two systems or a slight different effective annealing temperature for the two different types of samples. To exclude any artificial effect for the measured circular polarization, we have performed two complementary measurements. One is the evaluation of magnetic circular dichroism (MCD) effect in order to check the differential absorption of respective left and right circularly polarized light[20] through the CoFeB ferromagnetic layer in our sample. With linearly polarized excitation light, we have recorded the MCD signal by photoluminescence (PL) with different magnetic fields. As shown in Fig. 2(d), the MCD effect from the PMA spin-LED sample is lower than 1% in all investigated field range. The other measurement is to exclude artificial effects such as Zeeman splitting in the QW[10] by EL characterization of a reference sample without the CoFeB layer (replaced by a non-magnetic Ta layer in contact with MgO). As shown in Fig. 2(d), $P_C$ from the reference sample also shows less than 1% in all investigated field range. These two complementary measurements give a strong argument that the large $P_C$ we have observed is really due to the spin-polarized electron injected from the ultrathin CoFeB layer with PMA.

Another very interesting point is that the PMA property of our spin-injector can even persist up to room temperature. The inset of Fig. 3 shows the EL spectra with different circular polarizations at 300K under zero field. A clear difference of $I^{\sigma+}$ and $I^{\sigma-}$ components allows us to obtain $P_C$=8% at RT. The $P_C$ hysteresis loop is also in good agreement with the RT *M-H* hysteresis loop (Fig. 3). Although the out-of-plane coercivity $\mu_0 H_C$ is reduced at about 5 mT, it is sufficient to obtain an almost full remanent magnetization. As known in continuous wave EL experiments, the optical circular polarization $P_C$ is directly related to the injected electron spin polarization $P_E$ by a renormalization $F$ factor $P_C=F*P_E$ [33]. This $F$ factor which takes into account the electron spin relaxation mechanisms during its lifetime in the QW can be expressed by $F=1/(1+\tau/\tau_s)$, where $\tau$ and $\tau_s$ are respectively the electron lifetime and spin relaxation time in the QW[33,34]. In order to estimate $P_E$ at 300K, we have measured these characteristics times $\tau$ and $\tau_s$ at RT thanks to polarization and time-resolved photoluminescence (TRPL) measurements[35] on an identical bare *p-i-n* LED. We measure $\tau$~65 ps and $\tau_s$~45 ps at 300K, leading to an $F$ factor of 0.41. The corresponding injected electron spin polarization $P_E$ is thus ~20% at 300K.

Finally, we have measured $P_C$ as a function of the applied bias ($V_{bias}$) at 25K under zero field. As shown in Fig. 4, $P_C$ is found to be strongly dependent on the bias. The maximum EL circular polarization $P_C$ in remanence can even reach 20% at an optimal bias of 2.34V. The corresponding polarization-resolved EL spectra are shown in the inset of Fig. 4. $P_C$ decreases below and above this optimized bias. The origin of this behavior is still not completely understood at this step and will require further experiments. One possibility to explain the decrease of $P_C$ at low bias would be the complex behavior of the ratio $\tau/\tau_s$ as a function of the applied voltage (linked in particular to an increase of the carrier recombination time $\tau$ [9,25]). The decrease of $P_C$ at high bias could be also due to the dependence of $\tau/\tau_s$ as a function of $V_{bias}$, as well as to the D'Yakonov-Perel spin-relaxation mechanism for carriers injected with a large kinetic energy[15,36].

In conclusion, we have demonstrated a large EL circular polarization in (In)GaAs-based spin-LED at zero magnetic field with CoFeB/MgO perpendicular spin-injector. The value of $P_C$ at remanence is measured as large as 20% at 25K and still 8% at 300K, which is to our knowledge already five times higher (at low temperature) than the published results using any other PMA injectors[19-22,25]. Although the electric spin-injection efficiency $P_E$ is still lower than one of the best in-plane injector[15], we believe that detailed interfacial investigation and optimization of annealing effect could certainly lead to even larger improvement. This first demonstration of a robust and efficient remanent electrical spin injection thanks to an ultrathin CoFeB injector paves the way for future applications based on electrical control of circularly polarized light via STT in III-V opto-spintronic devices working at room temperature, as well as for devices where losses due to optical absorption are detrimental, as for example spin Vertical External Cavity Surface Emitting Lasers (VECSELS)[37] or spin Vertical Cavity Surface Emitting Lasers (VCSELS) where the magnetic injector would be embedded in the cavity.

**Acknowledgements**

We thank T. Hauet for the help for SQUID measurements and we also acknowledge Y. Zhen and D. Demaille for their help for TEM samples preparation. S. H. Liang acknowledges the joint PhD scholarship from China Scholarship Council (CSC). This work is supported by joint France-China (ANR-NSFC) SISTER project (ANR-11-IS10-0001) and ANR INSPIRE project (ANR-10-BLAN-1014).

**Figure captions:**

FIG.1. (Color online) (a) Schematic device structure of spin-LED. (b) HRTEM image of CoFeB/MgO PMA injector. Inset: low magnification image showing excellent homogeneity and low roughness of structures.

FIG.2. (Color online) (a) EL spectra at 25 K at zero magnetic field for $\sigma^+$ and $\sigma^-$ polarizations ($V_{bias}$=2.30V). (b) $P_C$ as a function of out-of-plane magnetic field measured at 25K, which is compared to the corresponding out-of-plane *M-H* hysteresis loop at 30K measured by SQUID. (c) *M-H* curves at 30K for spin-injector with 1.2 nm CoFeB and $T_{an}$=250°C for in-plane and out-of-plane configurations. (d) MCD induced by the ultrathin CoFeB electrode as a function of magnetic field at 25K, measured thanks to the luminescence circular polarization emitted under linearly polarized excitation of the spin-LED by a He-Ne laser. $P_C$ of EL for a reference sample without CoFeB layer (replaced by a non-magnetic Ta layer) as a function of magnetic field at 25K.

FIG.3. (Color online) (a) $P_C$ as a function of out-of-plane magnetic field measured at 300K ($V_{bias}$=2.30V), which is compared to the out-of-plane *M-H* hysteresis loop at 300K measured by SQUID. Inset: EL spectra at 300K with zero magnetic field for $\sigma^+$ and $\sigma^-$ polarizations.

FIG.4. (Color online) $P_C$ as a function of the applied bias at 25K. Inset: EL spectra at 25 K with zero magnetic field for $\sigma^+$ and $\sigma^-$ polarizations under a bias of 2.34V.

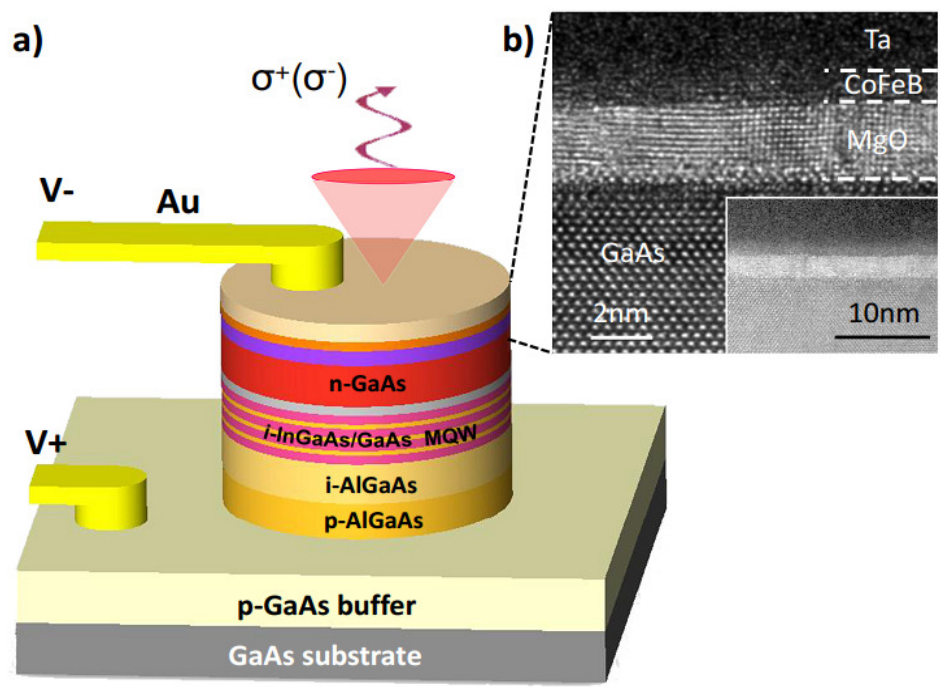

FIG.1

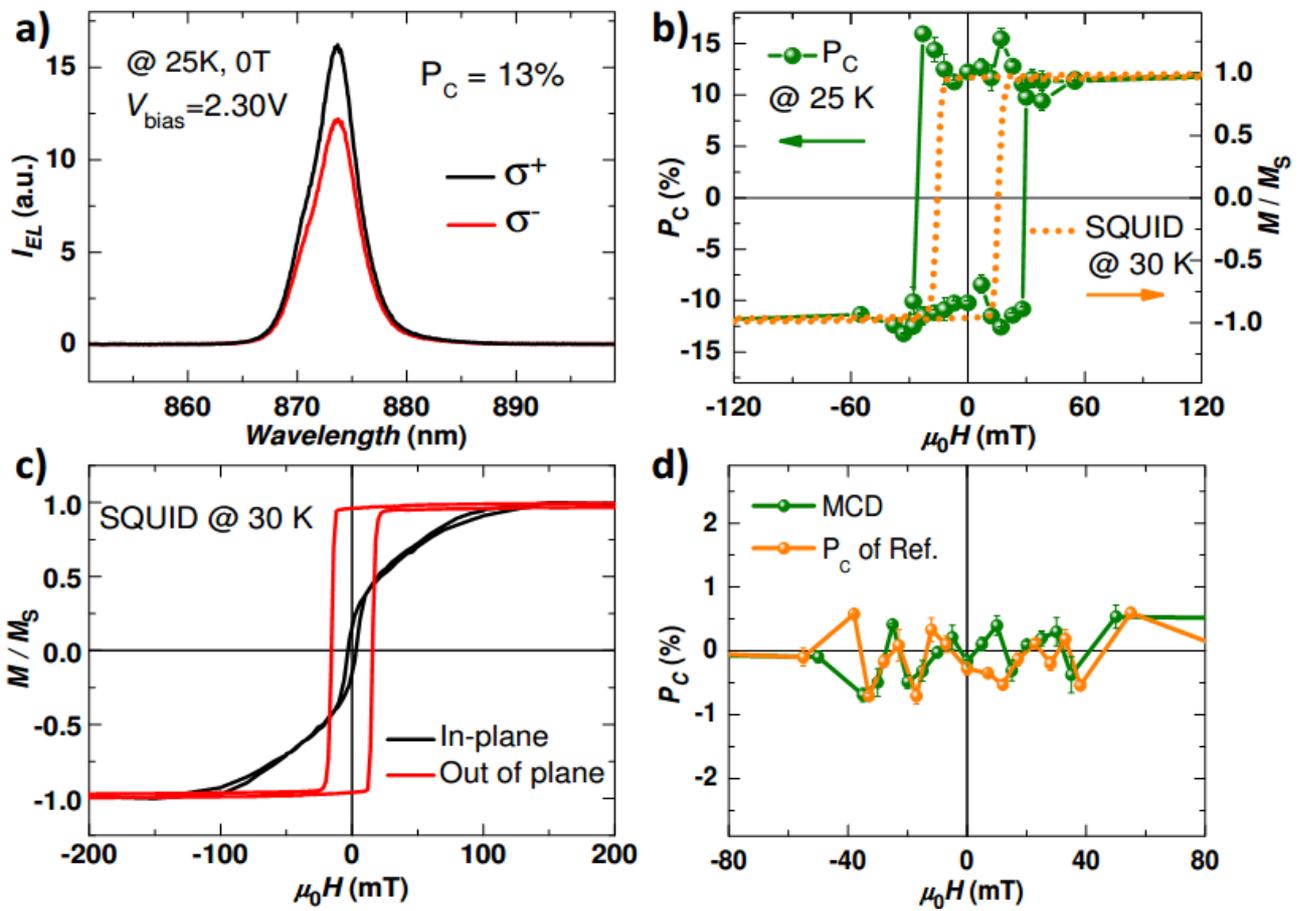

FIG.2

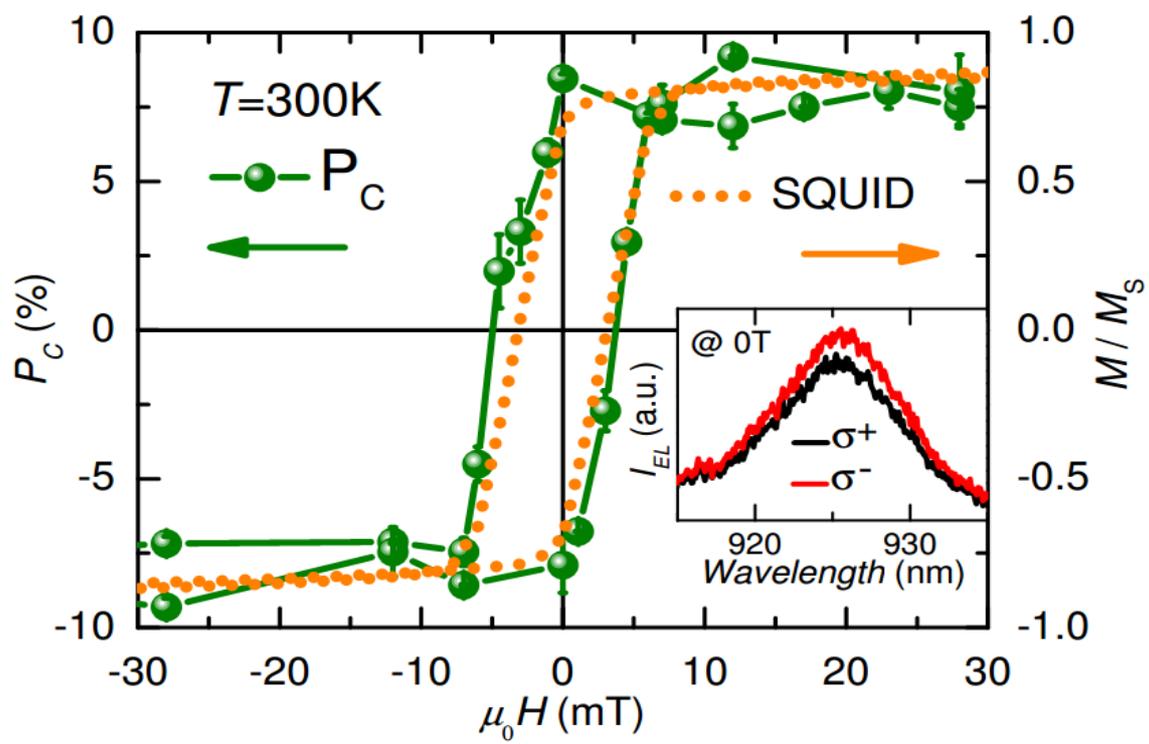

FIG.3

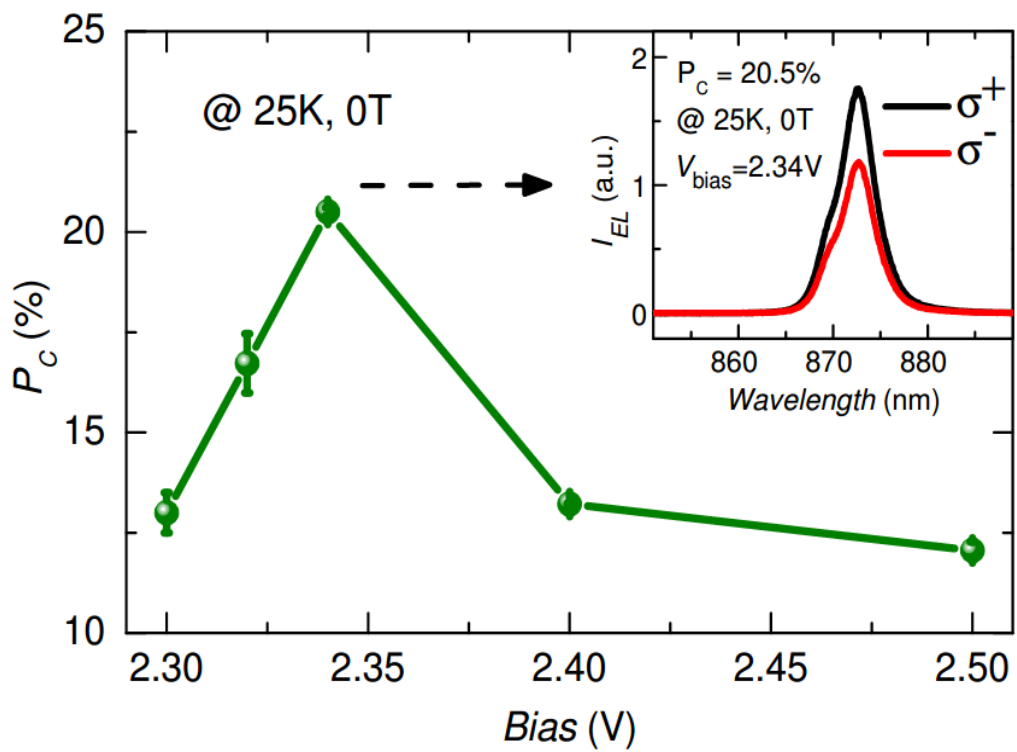

FIG.4